# Theory and Computation of the Spheroidal Wave Functions

P. E. Falloon, P. C. Abbott, and J. B. Wang

*School of Physics, The University of Western Australia*

*35 Stirling Hwy, Crawley WA 6009 AUSTRALIA.*

## Abstract

In this paper we report on a package, written in the *Mathematica* computer algebra system, which has been developed to compute the spheroidal wave functions of Meixner [J. Meixner and R.W. Schäfke, *Mathieusche Funktionen und Sphäroidfunktionen*, 1954] and is availlable online (www.physics.uwa.edu.au/~falloon/spheroidal/spheroidal.html). This package represents a substantial contribution to the existing software, since it computes the spheroidal wave functions to arbitrary precision for general complex parameters $\mu$, $\nu$, $\gamma$ and argument $z$; existing software can only handle integer $\mu$, $\nu$ and does not give arbitrary precision. The package also incorporates various special cases and computes analytic power series and asymptotic expansions in the parameter $\gamma$. The spheroidal wave functions of Flammer [C. Flammer, *Spheroidal Wave Functions*, 1957] are included as a special case of Meixner's more general functions. This paper presents a concise review of the general theory of spheroidal wave functions and a description of the formulas and algorithms used in their computation, and gives high-precision numerical examples.

PACS: `02.30.Gp`, `02.70.Wz`

## I. Introduction

Spheroidal wave functions are a class of special functions with many applications in physics and applied mathematics. They satisfy the differential equation

$$\frac{\mathrm{d}}{\mathrm{d}z}\left((1-z^2)\,\frac{\mathrm{d}f}{\mathrm{d}z}\right)+\left(\lambda_\nu^\mu(\gamma)+\gamma^2(1-z^2)-\frac{\mu^2}{1-z^2}\right)f(z)=0, \tag{1}$$

where $\mu$, $\nu$, $\gamma$ are arbitrary complex parameters. For many applications $\mu$, $\nu$ take on integer values, in which case they are denoted $m$, $n$. Although the general theory of spheroidal wave functions has been known for a long time [1-3], they are still regarded as difficult to compute, and at present there are few readily available computer packages available for their computation.

Solutions of Eq. (1) were first studied by Niven [4] in connection with a problem involving heat conduction in spheroidal bodies, and were subsequently investigated by a number of authors (see [5] and references therein). Early applications included the quantum mechanical two-centre problem and various electromagnetic boundary-value problems. The general theory and background on spheroidal wave functions is contained in the monograph by Flammer [2]. Other notable monographs are Stratton *et al.* [3], which has extensive tables of numerical values, and Komarov *et al.* [6]. These works focus exclusively on spheroidal wave functions with integer parameters $m$, $n$. Meixner developed the theory of spheroidal wave functions with arbitrary complex parameters $\mu$, $\nu$ (see [1, 7] and references therein).

Several packages have been developed recently to compute spheroidal wave functions: Thompson [8] (which uses an incorrect expansion for the angular functions of the second kind) and Li *et al.* [9] are two of the most recent. Both of these packages are only useful for small values of $\gamma$, do not provide arbitrary precision computation, and are limited to integer parameters $m$, $n$. Furthermore, no package currently in existence computes the power series and asymptotic expansions for the spheroidal wave functions.

The choice of a suitable notation and normalization for spheroidal wave functions presents a significant challenge, due to the large number of conventions in existence. The two main ones are those of Meixner [1] and Flammer [2] (also used in [10]). The latter is more commonly used in the literature, however its use is usually limited to integer parameters. Indeed, Flammer's normalization scheme cannot readily be generalized to noninteger parameters, because it normalizes functions to a different constant depending on the parity of $n-m$. Meixner's notation, though less commonly used, has the fundamental advantage that it is suitable for (and was in fact developed specifically to handle) the case of general complex parameters.

The purpose of this paper is to describe a package which has recently been developed [12] to compute spheroidal wave functions with arbitrary parameter values, which overcomes the shortcomings of existing software mentioned above. We have chosen to use the *Mathematica* computer algebra system [11], which is ideal due to its symbolic and high-precision numerical capabilities, as well as its large library of built-in special functions. We have taken a unique approach to the notation for the spheroidal functions, in that our package computes the functions of Flammer and Meixner as two distinct sets of functions. In this way, it is hoped that the package will be useful to as wide an audience as possible.

The layout of this paper is as follows. In Section II we present a concise review of the theory of the spheroidal wave functions, starting with their definition as series of Legendre and spherical Bessel functions. We then discuss the important special case for the angular functions of the second kind when $\mu + \nu$ is an integer, before discussing the spheroidal joining factor which relates the angular and radial functions. In Section III we describe the computation of the spheroidal wave functions, beginning with a discussion of the continued fraction and tridiagonal matrix methods used to compute the spheroidal eigenvalues. We then discuss the general numerical implementation of the spheroidal functions, and finally describe the approach used to generate the asymptotic and power series coefficients. We conclude with a discussion of some of the ways in which the accuracy of our numerical values can be verified. Four appendices are also included: Appendix A contains definitions for the Legendre and spherical Bessel functions which are used in the package; Appendix B contains formulas relating Flammer's spheroidal functions to those of Meixner; Appendix C contains a summary of some important mathematical identities satisfied by the spheroidal wave functions; finally, in Appendix D we present tables of sample function values to high-precision, for the purposes of comparison with other packages.

## II. Theory

### 2.1 The angular spheroidal wave functions

When $\gamma = 0$, Eq. (1) reduces to Legendre's differential equation

$$\frac{\mathrm{d}}{\mathrm{d}z}\left((1-z^2)\,\frac{\mathrm{d}f}{\mathrm{d}z}\right)+\left(\lambda_\nu^\mu(0)-\frac{\mu^2}{1-z^2}\right)f(z)=0. \tag{2}$$

The solutions to this equation are the (associated) Legendre functions of the first and second kind, $P_\nu^\mu(z)$ and $Q_\nu^\mu(z)$, with eigenvalue $\lambda_\nu^\mu(0) = \nu(\nu+1)$. Traditionally, these functions are defined differently depending on whether or not $z$ lies on the branch cut $z \in (-1, 1)$ (*e.g.* [10], §8.1, 8.3). For many applications—particularly those involving computer algebra—this convention is inconvenient, since it precludes the use of identities which are valid for all $z$. An alternative approach, which avoids this difficulty is to define two distinct types of Legendre function, each of which is valid for *all* values of $z$. The functions of Type I, $P_\nu^\mu(z)$ and $Q_\nu^\mu(z)$, are equivalent to those usually defined on the $(-1, 1)$ cut. The functions of Type II, which we denote $\mathfrak{P}_\nu^\mu(z)$ and $\mathfrak{Q}_\nu^\mu(z)$, are equivalent to the functions usually defined for $z \notin (-1, 1)$. In Appendix A we give the definitions and some important properties of these functions.

For nonzero $\gamma$, the angular spheroidal wave functions are defined as infinite series of the corresponding Legendre functions:

$$F_\nu^\mu(z;\gamma) = \sum_{k=-\infty}^{\infty} (-1)^k\, a_{\nu,k}^\mu(\gamma)\, f_{\nu+2k}^\mu(z). \tag{3}$$

Here we follow Meixner's notation so that $F$ = ps, qs, Ps, Qs and $f = P, Q, \mathfrak{P}, \mathfrak{Q}$ respectively.

It can readily be shown [1] that the series coefficients $a_{\nu,k}^\mu(\gamma)$ satisfy the three-term recurrence relation

$$A_{\nu,k}^\mu(\gamma)\, a_{\nu,k-1}^\mu(\gamma) + (B_{\nu,k}^\mu(\gamma) - \lambda_\nu^\mu(\gamma))\, a_{\nu,k}^\mu(\gamma) + C_{\nu,k}^\mu(\gamma)\, a_{\nu,k+1}^\mu(\gamma) = 0, \tag{4a}$$

where

$$\begin{aligned}
A_{\nu,k}^\mu(\gamma) &= -\gamma^2\, \frac{(\nu-\mu+2k-1)(\nu-\mu+2k)}{(2\nu+4k-3)(2\nu+4k-1)},\\
B_{\nu,k}^\mu(\gamma) &= (\nu+2k)(\nu+2k+1) - 2\gamma^2\, \frac{(\nu+2k)(\nu+2k+1)+\mu^2-1}{(2\nu+4k-1)(2\nu+4k+3)},\\
C_{\nu,k}^\mu(\gamma) &= -\gamma^2\, \frac{(\nu+\mu+2k+1)(\nu+\mu+2k+2)}{(2\nu+4k+3)(2\nu+4k+5)}.
\end{aligned} \tag{4b}$$

The series expansion in Eq. (3) is convergent only when the coefficients $a_{\nu,k}^\mu(\gamma)$ form a minimal solution to Eq. (4), (*i.e.* a solution with the property that $a_{\nu,k}^\mu(\gamma)/a_{\nu,k\mp1}^\mu(\gamma) \to 0$ as $k \to \pm\infty$ [16])—in which case it converges for all $z$. There is a countably infinite set of values for $\lambda$ that correspond to minimal solutions. The *spheroidal eigenvalue* $\lambda_\nu^\mu(\gamma)$ is defined as a function of $\mu$, $\nu$ and $\gamma$ by choosing the $\lambda$ value that reduces to $\nu(\nu+1)$ continuously as $\gamma \to 0$. In practice, this assignment is nontrivial and must be made numerically.

For integers $m$, $n$ with $n \geq |m|$, the coefficients $a^{m}_{n,k}(\gamma)$ are normalized so that

$$\int_{-1}^{1} \mathrm{ps}^{m}_{n}(t;\gamma)^2 \, \mathrm{d}t = \frac{2}{2\,n+1}\,\frac{(n+m)!}{(n-m)!},$$

which can be generalized in a natural way to give the relation for general $\mu$, $\nu$:

$$\sum_{k=-\infty}^{\infty} a^{\mu}_{\nu,k}(\gamma)^2 \,\frac{2\,\nu+1}{2\,\nu+4\,k+1}\,\frac{(\nu+\mu+1)_{2\,k}}{(\nu-\mu+1)_{2\,k}} = 1. \tag{5}$$

The sign of $a^{\mu}_{\nu,k}(\gamma)$ is determined by the condition that $\mathrm{ps}^{\mu}_{\nu}(z;\gamma) \to P^{\mu}_{\nu}(z)$ continuously as $\gamma \to 0$.

### 2.2 Angular functions of the second kind for integer $\mu+\nu$

The functions $Q^{\mu}_{\nu}(z)$ and $\mathbb{Q}^{\mu}_{\nu}(z)$ diverge when $\mu+\nu$ is a negative integer (Eq. A9), from which it immediately follows that

$$|\mathrm{qs}^{\mu}_{\nu}(z;\gamma)|,\ |\mathrm{Qs}^{\mu}_{\nu}(z;\gamma)| = \infty, \quad \mu+\nu = -1, -2, \ldots \tag{6}$$

For $\mu+\nu$ a non-negative integer, we have $C^{\mu}_{\nu,(-\mu-\nu-2+\delta)/2}(\gamma) = 0$, where $\delta = (\mu+\nu) \bmod 2$. From Eq. (4) we then find $a^{\mu}_{\nu,k}(\gamma) = 0$ for $k < -(\mu+\nu)/2$, and hence the series (6) becomes indeterminate for $\mathrm{qs}^{\mu}_{\nu}(z;\gamma)$ and $\mathrm{Qs}^{\mu}_{\nu}(z;\gamma)$, since the infinite basis functions are multiplied by zero series coefficients. The procedure for recovering a valid series representation in this case is reasonably straightforward, and is described (for integer parameters $m$, $n$) by Flammer [2].

The essential step is to use the transformation relations Eqs. (A10-11). Taking the case of $\mathbb{Q}^{\mu}_{\nu}(z)$ for definiteness, we substitute $\nu \to \nu+\epsilon$, where $\mu+\nu$ is an integer and $\epsilon \to 0$, into Eq. (A11) and immediately find

$$\sin(\epsilon\,\pi)\,\mathbb{Q}^{\mu}_{\nu+\epsilon}(z) = (-1)^{\mu+\nu}\,(\pi\,e^{\mathrm{i}\,\mu\,\pi}\cos((\nu+\epsilon)\,\pi)\,\mathbb{P}^{\mu}_{-\nu-\epsilon-1}(z) - \sin((\mu-\nu-\epsilon)\,\pi)\,\mathbb{Q}^{\mu}_{-\nu-\epsilon-1}(z)).$$

Multiplying by the series coefficient $a^{\mu}_{\nu+\epsilon,k}(\gamma)$ and taking the limit $\epsilon \to 0$ we have

$$\begin{aligned}&\lim_{\epsilon\to 0} a^{\mu}_{\nu+\epsilon,k}(\gamma)\,\mathbb{Q}^{\mu}_{\nu+2\,k+\epsilon}(z) = \\ &\qquad \tilde{a}^{\mu}_{\nu,k}(\gamma)\,(-1)^{\mu+\nu}\left(e^{\mathrm{i}\,\mu\,\pi}\cos(\nu\,\pi)\,\mathbb{P}^{\mu}_{-\nu-2\,k-1}(z) - \frac{1}{\pi}\sin((\mu-\nu)\,\pi)\,\mathbb{Q}^{\mu}_{-\nu-2\,k-1}(z)\right),\end{aligned} \tag{7}$$

where we define

$$\tilde{a}^{\mu}_{\nu,k}(\gamma) = \lim_{\epsilon\to 0}\frac{a^{\mu}_{\nu+\epsilon,k}(\gamma)}{\epsilon}, \quad k \leq k_0 - 1. \tag{8}$$

For $k \leq k_0 - 2$, the coefficients $\tilde{a}^{\mu}_{\nu,k}(\gamma)$ can be computed using

$$\frac{\tilde{a}^{\mu}_{\nu,k}(\gamma)}{\tilde{a}^{\mu}_{\nu,k+1}(\gamma)} = \frac{-C_k}{B_k - \lambda^{\mu}_{\nu}(\gamma) + A_k \,\frac{\tilde{a}^{\mu}_{\nu,k-1}(\gamma)}{\tilde{a}^{\mu}_{\nu,k}(\gamma)}}, \tag{9}$$

while for $k = k_0 - 1$, we have

$$\tilde{a}^{\mu}_{\nu,k_0-1}(\gamma) = \frac{-\tilde{C}^{\mu}_{\nu}(\gamma)\, a^{\mu}_{\nu,k_0}(\gamma)}{B_{k_0-1} - \lambda^{\mu}_{\nu}(\gamma) + A_{k_0-1} \,\frac{\tilde{a}^{\mu}_{\nu,k_0-2}(\gamma)}{\tilde{a}^{\mu}_{\nu,k_0-1}(\gamma)}}, \tag{10}$$

where

$$\tilde{C}^{\mu}_{\nu}(\gamma) = \lim_{\rho\to 0} \frac{C^{\mu}_{\nu+\rho,k_0-1}(\gamma)}{\rho} = \frac{(-1)^{\delta}\,\gamma^2}{(2\,\mu - 2\,\delta - 1)\,(2\,\mu - 2\,\delta + 1)}. \tag{11}$$

For $\mu + \nu = 0,\ 1,\ 2,\ \ldots$ the series for $\mathrm{Qs}^{\mu}_{\nu}(z;\gamma)$ therefore reads

$$\begin{aligned}\mathrm{Qs}^{\mu}_{\nu}(z;\gamma) = \sum_{k=k_0}^{\infty} (-1)^k\, a^{\mu}_{\nu,k}(\gamma)\, \mathbb{Q}^{\mu}_{\nu+2\,k}(z) + (-1)^{\mu+\nu} \sum_{k=-\infty}^{k_0-1} (-1)^k\, \tilde{a}^{\mu}_{\nu,k}(\gamma) \\ \times\Big(e^{i\,\mu\,\pi}\cos(\nu\,\pi)\,\mathbb{P}^{\mu}_{-\nu-2\,k-1}(z) - \frac{\sin((\mu-\nu)\,\pi)}{\pi}\,\mathbb{Q}^{\mu}_{-\nu-2\,k-1}(z)\Big).\end{aligned} \tag{12a}$$

For $\mathrm{qs}^{\mu}_{\nu}(z;\gamma)$ we follow a directly anagolous argument starting from Eq. (A10) in place of Eq. (A11), and obtain

$$\begin{aligned}\mathrm{qs}^{\mu}_{\nu}(z;\gamma) = \sum_{k=k_0}^{\infty} (-1)^k\, a^{\mu}_{\nu,k}(\gamma)\, Q^{\mu}_{\nu+2\,k}(z) + (-1)^{\mu+\nu} \sum_{k=-\infty}^{k_0-1} (-1)^k\, \tilde{a}^{\mu}_{\nu,k}(\gamma) \\ \times\Big(\cos(\mu\,\pi)\cos(\nu\,\pi)\,P^{\mu}_{-\nu-2\,k-1}(z) - \frac{\sin((\mu-\nu)\,\pi)}{\pi}\,Q^{\mu}_{-\nu-2\,k-1}(z)\Big).\end{aligned} \tag{12b}$$

For integers $m$, $n$ these expressions reduce to

$$\mathrm{Qs}^{m}_{n}(z;\gamma) = \sum_{k=k_0}^{\infty} (-1)^k\, a^{m}_{n,k}(\gamma)\, \mathbb{Q}^{m}_{n+2\,k}(z) + \sum_{k=-\infty}^{k_0-1} (-1)^k\, \tilde{a}^{m}_{n,k}(\gamma)\, \mathbb{P}^{m}_{-n-2\,k-1}(z), \tag{13a}$$

$$\mathrm{qs}^{m}_{n}(z;\gamma) = \sum_{k=k_0}^{\infty} (-1)^k\, a^{m}_{n,k}(\gamma)\, Q^{m}_{n+2\,k}(z) + \sum_{k=-\infty}^{k_0-1} (-1)^k\, \tilde{a}^{m}_{n,k}(\gamma)\, P^{m}_{-n-2\,k-1}(z). \tag{13b}$$

### 2.3 The radial spheroidal wave functions

In the limit $\gamma \to 0$ and $z \to \infty$ such that $\gamma\, z = \text{constant}$, the two regular singularities of Eq. (1) (at $z = \pm 1$) coalesce. This can be seen by changing variables to $\zeta = \gamma\, z$, substituting $f(z) = (1 - 1/z^2)^{\mu/2}\, g(\zeta)$, and letting $\gamma \to 0$. Eq. (1) then becomes

$$\zeta^2 \frac{\mathrm{d}^2 g}{\mathrm{d}\zeta^2} + 2\,\zeta \frac{\mathrm{d}g}{\mathrm{d}\zeta} + (\zeta^2 - \lambda)\, g(\zeta) = 0, \tag{14}$$

which is satisfied by the spherical Bessel functions $j_\nu(\zeta)$ and $y_\nu(\zeta)$ ([10], Ch.10 and Appendix A). The radial spheroidal wave functions are defined in terms of these by

$$S_\nu^{\mu\,(k)}(z;\gamma) = \frac{(1 - 1/z^2)^{\mu/2}}{A_\nu^{-\mu}(\gamma)} \sum_{k=-\infty}^{\infty} a_{\nu,k}^{-\mu}(\gamma)\, f_{\nu+2k}(\gamma\, z), \tag{15}$$

where $k = 1, 2$ and $f = j, y$ respectively, and

$$A_\nu^\mu(\gamma) = \sum_{k=-\infty}^{\infty} (-1)^k\, a_{\nu,k}^\mu(\gamma). \tag{16}$$

It has been shown [13] that the functions defined in Eq. (15) are indeed solutions to Eq. (1) which are absolutely convergent for $|z| > 1$. The normalization factor $A_\nu^\mu(\gamma)$ is chosen so that the following limits are satisfied:

$$S_\nu^{\mu\,(1)}(z;\gamma) \underset{\gamma z\to\infty}{\longrightarrow} \frac{1}{\gamma\, z} \sin\!\left(\gamma\, z - \frac{\nu\,\pi}{2}\right), \quad S_\nu^{\mu\,(2)}(z;\gamma) \underset{\gamma z\to\infty}{\longrightarrow} -\frac{1}{\gamma\, z} \cos\!\left(\gamma\, z - \frac{\nu\,\pi}{2}\right). \tag{17}$$

The functions $S_\nu^{\mu\,(k)}(z;\gamma)$ have branch cuts in the complex $z$-plane along the line $(-1/\gamma\,\infty, 0)$, for $\nu \notin \mathbb{Z}$, and on the interval $(-1, 1)$, for $\mu/2 \notin \mathbb{Z}$.

## 2.4 Joining relations between angular and radial functions

The angular and radial spheroidal wave functions can both be considered as functions over the entire complex $z$-plane. From a computational point of view, however, their series expansions are only useful over a restricted subset of the complex plane. For the angular functions, the series (3) is convergent over the entire $z$-plane, but for $|z| > 1$ it becomes too slowly convergent to be of any practical use. For the radial functions the situation is even worse—the series (15) is in general not convergent inside the unit circle $|z| < 1$. To allow computation of the functions over the entire complex plane, Meixner and Schäfke [1] introduced a *joining factor* that relates the angular and radial functions.

Using well-known series expansions for $\mathfrak{Q}_\nu^\mu(z)$ and $j_\nu(z)$, it is possible to find the following series expansions for $\mathrm{Qs}_\nu^\mu(z;\gamma)$ and $S_\nu^{\mu\,(1)}(z;\gamma)$ [12]:

$$\mathrm{Qs}_{\nu}^{\mu}(z;\gamma) = 2^{-\nu-1}\,\sqrt{\pi}\,e^{\mathrm{i}\,\mu\,\pi}\,z^{-\mu-\nu-1}\,(z-1)^{\mu/2}\,(z+1)^{\mu/2}$$
$$\times\sum_{j-\infty}^{\infty}\frac{1}{(2\,z)^{2\,j}}\sum_{k=0}^{\infty}\frac{(-1)^{j-k}\,\Gamma(2\,j+\mu+\nu+1)}{k!\,\Gamma(2\,j-k+\nu+\frac{3}{2})}\,a_{\nu,2\,j-2\,k}^{\mu}(\gamma), \tag{18a}$$

$$S_{\nu}^{\mu\,(1)}(z;\gamma) = \frac{\sqrt{\pi}\,(1-1/z^2)^{\mu/2}\,(\gamma\,z)^{\nu}}{2^{\nu+1}\,A_{\nu}^{-\mu}(\gamma)}\sum_{j=-\infty}^{\infty}\frac{1}{(2\,z)^{2\,j}}\sum_{k=0}^{\infty}\frac{(-1)^k\,2^{4\,j}\,a_{\nu,-2\,j-2\,k}^{-\mu}(\gamma)}{\Gamma(-2\,j-k+\nu+\frac{3}{2})\,k!\,\gamma^{2\,j}}. \tag{18b}$$

Comparison of Eqs. (18a) and (18b) reveals that the expansions of $\mathrm{Qs}_{-\nu-1}^{\mu}(z;\gamma)$ and $S_{\nu}^{\mu\,(1)}(z;\gamma)$ involve identical powers of $z$. Furthermore, it is obvious from Eq. (A13) that the expansion for $S_{\nu}^{\mu\,(2)}(z;\gamma)$ will *not* involve the same powers of $z$. Now, $\mathrm{Qs}_{-\nu-1}^{\mu}(z;\gamma)$ must be expressible as a linear combination of the functions $S_{\nu}^{\mu\,(1,2)}(z;\gamma)$, since it satisfies the same differential equation, so it follows that $\mathrm{Qs}_{-\nu-1}^{\mu}(z;\gamma)$ and $S_{\nu}^{\mu\,(1)}(z;\gamma)$ must be equal up to a constant factor. Comparing (18a) and (18b), we have then that the ratio

$$\left(\sum_{k=0}^{\infty}\frac{(-1)^k\,2^{4\,j}\,a_{\nu,-2\,j-2\,k}^{-\mu}(\gamma)}{\Gamma(-2\,j-k+\nu+\frac{3}{2})\,k!\,\gamma^{2\,j}}\right)\Bigg/\left(\sum_{k=0}^{\infty}\frac{(-1)^{j-k}\,\Gamma(2\,j+\mu-\nu)\,a_{\nu,-2\,j+2\,k}^{\mu}(\gamma)}{\Gamma(2\,j-k-\nu+\frac{1}{2})\,k!}\right)$$

must be independent of $j$.

In light of the above result, we define the *spheroidal joining factor* $K_{\nu}^{\mu}(\gamma)$ by the relation

$$S_{\nu}^{\mu\,(1)}(z;\gamma) = K_{\nu}^{\mu}(\gamma)\,\frac{\sin((\mu-\nu)\,\pi)}{\pi}\,e^{-\mathrm{i}\,(\mu+\nu)\,\pi}\,\frac{(1-1/z^2)^{\mu/2}\,(\gamma\,z)^{\nu}}{\gamma^{\nu}\,z^{\nu-\mu}\,(z-1)^{\mu/2}\,(z+1)^{\mu/2}}\,\mathrm{Qs}_{-\nu-1}^{\mu}(z;\gamma). \tag{19}$$

The trigonometric and exponential factors are included so that the joining factor relations reduce to a simple form for integers $m$, $n$. Note also the factor involving various powers of $z$, which includes the branch cut information for the two functions. Comparing terms in Eqs. (18a) and (18b) for any particular $j$ we can obtain an explicit definition for $K_{\nu}^{\mu}(\gamma)$. For definiteness we choose $j = 0$ and obtain

$$K_{\nu}^{\mu}(\gamma) = e^{\mathrm{i}\,\nu\,\pi}\,2^{-2\,\nu-1}\,\Gamma(\nu-\mu+1)\,\frac{\gamma^{\nu}}{A_{\nu}^{-\mu}(\gamma)}$$
$$\times\left(\sum_{k=0}^{\infty}\frac{(-1)^k\,a_{\nu,-2\,k}^{-\mu}(\gamma)}{\Gamma(-k+\nu+\frac{3}{2})\,k!}\right)\Bigg/\left(\sum_{k=0}^{\infty}\frac{a_{\nu,2\,k}^{\mu}(\gamma)}{\Gamma(-k-\nu+\frac{1}{2})\,k!}\right). \tag{20}$$

Using this equation and the symmetry relations given in Appendix C it is possible to obtain

$$A_{\nu}^{-\mu}(\gamma)\,K_{-\nu-1}^{\mu}(\gamma)\,A_{\nu}^{\mu}(\gamma)\,K_{\nu}^{-\mu}(\gamma) = \frac{\pi}{\gamma\sin((\mu+\nu)\,\pi)}, \tag{21}$$

which is useful in constructing joining relations for integers $m$, $n$.

## III. Numerical computation

### 3.1 The spheroidal eigenvalues $\lambda_\nu^\mu(\gamma)$

As we mentioned in Section 2.1, the spheroidal eigenvalues $\lambda_\nu^\mu(\gamma)$ are minimal solutions of the three-term recurrence (4). There are two standard procedures for finding such solutions. The first was developed independently by Bouwkamp [14] and Blanch [15], and makes use of a fundamental equivalence between three-term recurrences and continued fractions [16]. This provides a method for determining the eigenvalues numerically to high precision, although it relies on the availability of a sufficiently accurate starting estimate for the eigenvalue. The second method, due to Hodge [17], involves expressing the three-term recurrence as an infinite tridiagonal matrix equation. It is complementary to the first in the sense that it provides an excellent method for generating accurate starting estimates for the eigenvalues, but is inefficient for obtaining high precision eigenvalues. We now discuss both methods in turn.

*Continued fraction method*

Defining

$$\begin{aligned} \alpha_{\nu,k}^\mu(\gamma) &= A_{\nu,k}^\mu(\gamma)\, C_{\nu,k-1}^\mu(\gamma), \\ \beta_{\nu,k}^\mu(\gamma) &= B_{\nu,k}^\mu(\gamma), \\ N_{\nu,k}^\mu(\gamma) &= C_{\nu,k-1}^\mu(\gamma)\, \frac{a_{\nu,k}^\mu(\gamma)}{a_{\nu,k-1}^\mu(\gamma)}, \end{aligned} \tag{22}$$

the three-term recurrence (4a) can be rewritten in ascending and descending form as

$$N_{\nu,k+1}^\mu(\gamma) = \frac{\alpha_{\nu,k+1}^\mu(\gamma)}{\beta_{\nu,k+1}^\mu(\gamma) - \lambda_\nu^\mu(\gamma) - N_{\nu,k+2}^\mu(\gamma)}, \quad N_{\nu,k+1}^\mu(\gamma) = \beta_{\nu,k}^\mu(\gamma) - \lambda_\nu^\mu(\gamma) - \frac{\alpha_{\nu,k}^\mu(\gamma)}{N_{\nu,k}^\mu(\gamma)}.$$

Setting $k = 0$ and iterating these relations we obtain

$$\mathcal{U}_\nu^{\mu(1)}(\gamma, \lambda) + \mathcal{U}_\nu^{\mu(2)}(\gamma, \lambda) = 0, \tag{23a}$$

where we have defined

$$\mathcal{U}_\nu^{\mu\,(1)}(\gamma,\lambda) = \beta_0 - \lambda - \frac{\alpha_0}{\beta_{-1} - \lambda -}\,\frac{\alpha_{-1}}{\beta_{-2} - \lambda -}\cdots$$
$$\mathcal{U}_\nu^{\mu\,(2)}(\gamma,\lambda) = -\frac{\alpha_1}{\beta_1 - \lambda -}\,\frac{\alpha_2}{\beta_2 - \lambda -}\,\frac{\alpha_3}{\beta_3 - \lambda -}\cdots \tag{23b}$$

Here we are using a standard notational convention for continued fractions [18]:

$$\frac{a_1}{b_1 +}\,\frac{a_2}{b_2 +}\,\frac{a_3}{b_3 +}\cdots = \frac{a_1}{b_1 + \frac{a_2}{b_2 + \frac{a_3}{b_3 + \cdots}}}\,.$$

Eq. (23a) is a transcendental equation in $\lambda$, whose roots are the spheroidal eigenvalues $\lambda_{\nu+2k}^{\mu}(\gamma)$. The method of Bouwkamp and Blanch consists of differentiating the left side of Eq. (23a) and using Newton's method. Differentiating Eq. (23b) with respect to $\lambda$ we obtain

$$\frac{\partial \mathcal{U}_\nu^{\mu\,(1)}(\gamma,\lambda)}{\partial\lambda} = -\left(1 + \frac{\alpha_0}{N_0^2} + \frac{\alpha_0}{N_0^2}\,\frac{\alpha_{-1}}{N_{-1}^2} + \frac{\alpha_0}{N_0^2}\,\frac{\alpha_{-1}}{N_{-1}^2}\,\frac{\alpha_{-2}}{N_{-2}^2} + \cdots\right),$$
$$\frac{\partial \mathcal{U}_\nu^{\mu\,(2)}(\gamma,\lambda)}{\partial\lambda} = -\left(\frac{N_1^2}{\alpha_1} + \frac{N_1^2}{\alpha_1}\,\frac{N_2^2}{\alpha_2} + \frac{N_1^2}{\alpha_1}\,\frac{N_2^2}{\alpha_2}\,\frac{N_3^2}{\alpha_3} + \cdots\right). \tag{24}$$

To apply Newton's method to Eq. (23) we begin with a starting estimate $\lambda_0$ and iterate Newton's formula,

$$\delta\lambda_i = -\frac{\mathcal{U}_\nu^{\mu\,(1)}(\gamma,\lambda_{i-1}) + \mathcal{U}_\nu^{\mu\,(2)}(\gamma,\lambda_{i-1})}{\frac{\partial \mathcal{U}_\nu^{\mu\,(1)}(\gamma,\lambda_{i-1})}{\partial\lambda} + \frac{\partial \mathcal{U}_\nu^{\mu\,(2)}(\gamma,\lambda_{i-1})}{\partial\lambda}}, \tag{25}$$

starting with $i = 1$, until the size of the $i$th iterate $\delta\lambda_i$ is beneath the desired level of precision. Provided sufficient precision is used throughout the application of this algorithm, it can be used to obtain $\lambda_\nu^\mu(\gamma)$ to arbitrarily high accuracy. Usually the intermediate working precision needs to be significantly higher than the desired final precision, as precision is almost always lost in the process of numerical computation.

### *Tridiagonal matrix method*

In order to use the continued fraction method just described, it is necessary to begin with a reasonably accurate starting value. Traditionally, most authors have used a power series expansion of $\lambda_\nu^\mu(\gamma)$ in powers of $\gamma$ [2], which has a very small radius of convergence (approximately between 4 and 5). Hodge [17] appears to have been the first to solve the recurrence Eq. (4) by recasting it as a tridiagonal matrix equation:

$$\begin{pmatrix} \cdot & \cdot & & & & & \\ \cdot & \cdot & \cdot & & & & \\ & A_{-2} & B_{-2} & C_{-2} & & & \\ & & A_0 & B_0 & C_0 & & \\ & & & A_2 & B_2 & C_2 & \\ & & & & \cdot & \cdot & \cdot \\ & & & & & \cdot & \cdot \end{pmatrix} \begin{pmatrix} \cdot \\ \cdot \\ a_{-2} \\ a_0 \\ a_2 \\ \cdot \\ \cdot \end{pmatrix} = \lambda_\nu^\mu(\gamma) \begin{pmatrix} \cdot \\ \cdot \\ a_{-2} \\ a_0 \\ a_2 \\ \cdot \\ \cdot \end{pmatrix}. \tag{26}$$

Truncating this equation in both directions yields a finite matrix, the eigenvalues of which will be approximations to the actual eigenvalues $\lambda_\nu^\mu(\gamma)$. Since eigenvalues of tridiagonal matrices can be evaluated very efficiently, this method allows approximate eigenvalues to be generated rapidly. These matrix eigenvalues are ideal starting estimates for the continued fraction method described above. Surprisingly, to our knowledge no author seems to have previously used this hybrid approach.

One issue which presents a practical challenge to the computation of the spheroidal eigenvalues is actually deciding which matrix eigenvalue corresponds to which value of $\nu$: all that can be said in general is that each eigenvalue of the matrix corresponds to $\lambda_{\nu+2k}^\mu(\gamma)$ for some integer $k$. When $\gamma^2$ is real and $m$, $n$ are integers, the eigenvalues are strictly ordered and hence the correspondence is trivial. However, in the complex $\gamma$ plane the eigenvalues have a complicated branch cut structure, and the ordering relation does not apply. The same is true for non-integer $\mu, \nu$. One rather cumbersome solution to the problem is to start with an eigenvalue $\lambda_n^m(\gamma_0)$ for which the ordering relation does hold (*i.e.* $m$, $n$ are integers $\gamma_0$ is on the real or imaginary axis), and then follow a curve in the $\mu, \nu, \gamma$ parameter space towards the desired eigenvalue $\lambda_\nu^\mu(\gamma)$. Provided the step sizes are small enough, each eigenvalue can be used to choose the correct starting value from the matrix at the subsequent step. However, this approach leads to an intractable amount of computation unless parameter values are very small. In the package, therefore, we simply include an optional extra argument to the eigenvalue function, which allows a starting estimate to be given by the user. Using this, it is straightforward to implement the above iterative procedure for particular cases.

### 3.2 The spheroidal wave functions

Once the eigenvalue $\lambda_\nu^\mu(\gamma)$ has been found, it is relatively straightforward to compute the actual spheroidal wave functions using Eqs. (3) and (15). There are two main parts to this computation: generating the series coefficients $a_{\nu,k}^\mu(\gamma)$ and generating the basis functions (Legendre or spherical Bessel functions). We now discuss each of these in turn.

*Series coefficients*

To generate the series coefficients $a^{\mu}_{\nu,k}(\gamma)$ we follow the standard approach of rewriting the recurrence relation (4) in terms of ascending and descending ratios:

$$\begin{aligned}\frac{a^{\mu}_{\nu,k}(\gamma)}{a^{\mu}_{\nu,k+1}(\gamma)} &= \frac{-C^{\mu}_{\nu,k}(\gamma)}{B^{\mu}_{\nu,k}(\gamma) - \lambda^{\mu}_{\nu}(\gamma) + A^{\mu}_{\nu,k}(\gamma)\,\frac{a^{\mu}_{\nu,k-1}(\gamma)}{a^{\mu}_{\nu,k}(\gamma)}},\\ \frac{a^{\mu}_{\nu,k}(\gamma)}{a^{\mu}_{\nu,k-1}(\gamma)} &= \frac{-A^{\mu}_{\nu,k}(\gamma)}{B^{\mu}_{\nu,k}(\gamma) - \lambda^{\mu}_{\nu}(\gamma) + C^{\mu}_{\nu,k}(\gamma)\,\frac{a^{\mu}_{\nu,k+1}(\gamma)}{a^{\mu}_{\nu,k}(\gamma)}}.\end{aligned} \tag{27}$$

These ratios converge like $1/k^2$ as $k \to \pm\infty$, so we can set them to approximately to zero for some large value $|k| = k_{\max}$. We then iterate Eq. (27) from $k = \pm k_{\max}$ to $k = 0$ to obtain $a^{\mu}_{\nu,k}(\gamma)/a^{\mu}_{\nu,0}(\gamma)$ for $-k_{\max} \le k \le k_{\max}$. Once again, in this recursive process precision is usually lost with each step, and hence it is necessary to work with an intermediate precision substantially greater than the final desired precision. In practice, our package begins with a working precision of 100 extra digits, then tests at the end whether the final precision is high enough. If it is not, the working precision is increased by 100 and the coefficients computed again.

The normalization relation (5) can be rewritten in the form

$$a^{\mu}_{\nu,0}(\gamma) \simeq \left( \sum_{k=-k_{\max}}^{k_{\max}} \left( \frac{a^{\mu}_{\nu,k}(\gamma)}{a^{\mu}_{\nu,0}(\gamma)} \right)^2 \frac{2\nu+1}{2\nu+2k+1} \frac{(\nu+\mu+1)_k}{(\nu-\mu+1)_k} \right)^{-1/2},$$

allowing us to obtain $a^{\mu}_{\nu,0}(\gamma)$ and hence the correctly normalized list of coefficients $a^{\mu}_{\nu,k}(\gamma)$ for $-k_{\max} \le k \le k_{\max}$.

The question is now whether or not the chosen value of $k_{\max}$ is large enough that the omitted "tail" of the series is insignificant to the given level of precision. To answer this question rigorously is not easy, and in the package we settle on simply checking that the magnitude of $a^{\mu}_{\nu,\pm k}(\gamma)$ is negligible compared to the largest series coefficient, *i.e.*

$$\frac{|a^{\mu}_{\nu,\pm k_{\max}}(\gamma)|}{\max(\{|a^{\mu}_{\nu,k}(\gamma)| \mid -k_{\max} \le k \le k_{\max}\})} < 10^{-\text{prec}}.$$

Although not mathematically rigorous, extensive numerical testing shows that this is an eminently reasonable condition.

### *Basis functions*

In principle, the basis functions could be computed using *Mathematica*'s built-in functions. However, it is much more efficient to start with the basis functions with $k = 0, 1$ and then use recurrence relations satisfied by the basis functions to generate the basis functions for all other values of $k$. The required relations are

$$\begin{gathered}\frac{(\mu-\nu-2)(\mu-\nu-1)}{(2\nu+1)(2\nu+3)} f^{\mu}_{\nu+2}(z) \\ +\left(\frac{(2\nu(\nu+1)-2\mu^2-1)}{(2\nu-1)(2\nu+3)} - z^2\right) f^{\mu}_{\nu}(z) + \frac{(\mu+\nu-1)(\mu+\nu)}{(2\nu-1)(2\nu+1)} f^{\mu}_{\nu-2}(z) = 0,\end{gathered} \tag{28}$$

for $f = \mathsf{P}, \mathsf{Q}, P, Q$, and

$$\frac{f_{\nu-2}(z)}{2\nu-1} + (2\nu+1)\left(\frac{2}{(2\nu-1)(2\nu+3)} - \frac{1}{z^2}\right) f_{\nu}(z) + \frac{f_{\nu+2}(z)}{2\nu+3} = 0, \tag{29}$$

for $f = j, y$. These relations are readily obtained from the standard recurrence formulas §8.5.3 and §10.1.19 in [10]. Once again, these computations must be carried out at a much higher precision than is required at the end.

## 3.3 Series expansions

The eigenvalues $\lambda^{\mu}_{\nu}(\gamma)$ and the ratios $a^{\mu}_{\nu,k}(\gamma)/a^{\mu}_{\nu,0}(\gamma)$ can be expanded in powers of $\gamma^2$ [2], leading to approximations which are useful for $|\gamma| \lesssim 4$-$5$. However, for these values of $\gamma$ the tridiagonal matrix approach of Section 3.1 is much more useful for obtaining numerical values, so in practice the power series is mainly of theoretical interest. We now describe how the power series expansions are computed in our package.

We begin with the following *ansätze*:

$$\begin{aligned}\lambda^{\mu}_{\nu}(\gamma) &= \sum_{j=0}^{\infty} \ell^{\mu\nu}_{j}\, \gamma^{2j}, \\ \frac{a^{\mu}_{\nu,k}(\gamma)}{a^{\mu}_{\nu,0}(\gamma)} &= \sum_{j=0}^{\infty} \alpha^{\mu\nu}_{jk}\, \gamma^{2j}, \quad k = 0, \pm 1, \pm 2, \ldots\end{aligned} \tag{30}$$

By inspection we immediately have

$$\alpha^{\mu\nu}_{0,0} = 1, \quad \alpha^{\mu\nu}_{j,0} = \alpha^{\mu\nu}_{0,k} = 0, \text{ for } j, k \neq 0. \tag{31}$$

Substituting the expansions in (30) into the recurrence relation (4) with $k = 0$, and defining

$$
\begin{aligned}
A_{\mu\nu k} &= \frac{(\mu+\nu+k+1)\,(\mu+\nu+k+2)}{(2\,\nu+2\,k+3)\,(2\,\nu+2\,k+5)}, \\
B_{\mu\nu k} &= \frac{1}{2}\left(1-\frac{4\,\mu^2-1}{(2\,\nu+2\,k-1)\,(2\,\nu+2\,k+3)}\right), \\
C_{\mu\nu k} &= \frac{(\nu-\mu+k)\,(\nu-\mu+k-1)}{(2\,\nu+2\,k-3)\,(2\,\nu+2\,k-1)},
\end{aligned}
$$

we find

$$
\frac{\nu\,(\nu+1)-\ell_0^{\mu\nu}}{\gamma^2}+\beta_{\mu,\nu,0}-\ell_1^{\mu\nu}+\sum_{j=1}^{\infty}(A_{\mu,\nu,0}\,\alpha_{j,1}^{\mu\nu}+C_{\mu,\nu,0}\,\alpha_{j,-1}^{\mu\nu}-\ell_{j+1}^{\mu\nu})\,\gamma^{2\,j}=0. \tag{32}
$$

Since this is true for all values of $\gamma$ it follows that

$$
\begin{aligned}
\ell_0^{\mu\nu} &= \nu\,(\nu+1), \\
\ell_1^{\mu\nu} &= B_{\mu,\nu,0} = \frac{1}{2}\left(1-\frac{4\,\mu^2-1}{(2\,\nu-1)\,(2\,\nu+3)}\right), \\
\ell_j^{\mu\nu} &= A_{\mu,\nu,0}\,\alpha_{j-1,1}^{\mu\nu}+C_{\mu,\nu,0}\,\alpha_{j-1,-1}^{\mu\nu}, \qquad j=2,\,3,\,\ldots
\end{aligned} \tag{33}
$$

This provides a recursive method for determining $\ell_j^{\mu\nu}$ when the coefficients $\alpha_{j-1,\pm 1}^{\mu\nu}$ are known. To find these coefficients we substitute Eq. (30) into Eq. (4) for general $k$. After some manipulation we find

$$
\begin{aligned}
\alpha_{jk}^{\mu\nu} &= \frac{1}{(\nu+k)\,(\nu+k+1)-\ell_0^{\mu\nu}} \\
&\quad\times\left(\sum_{i=0}^{j-1}\ell_{i+1}^{\mu\nu}\,\alpha_{j-i-1,k}^{\mu\nu}-(A_{\mu\nu k}\,\alpha_{j-1,k+2}^{\mu\nu}+B_{\mu\nu k}\,\alpha_{j-1,k}^{\mu\nu}+C_{\mu\nu k}\,\alpha_{j-1,k-2}^{\mu\nu})\right).
\end{aligned} \tag{34}
$$

Now let $k$ be a positive integer and suppose that $\alpha_{j-1,k-2}^{\mu\nu}=0$ and $\alpha_{rs}^{\mu\nu}=0$ for all $r \le j-1$ and $s>k$. Then Eq. (34) shows that $\alpha_{jk}^{\mu\nu}=0$. This fact, together with the initial conditions $\alpha_{0,k}^{\mu\nu}=0$ for $k=\pm 2,\,\pm 4,\,\ldots$, proves by induction that

$$
\alpha_{jk}^{\mu\nu}=0, \quad j<|k|/2, \tag{35}
$$

and hence $a_{\nu,\pm k}^{\mu}(\gamma)/a_{\nu,0}^{\mu}(\gamma)=O(\gamma^{2\,k})$ for $k=0,\,1,\,2,\,\ldots$ This result could also have been deduced directly from the recurrence Eq. (4). Eqs. (33-35) constitute the recursive scheme by which we compute the power series expansions (30a-b).

The key to an efficient implementation of recursive algorithms of this kind is to use *dynamic programming*, whereby coefficients are "cached" once they are generated, allowing subsequent coefficients to be generated more quickly. In *Mathematica*, this is achieved with a definition of the form $f$(x_) := $f(x)$ = …

With the expansions for the ratios $a^{\mu}_{\nu,k}(\gamma)/a^{\mu}_{\nu,0}(\gamma)$ computed, it is not difficult to obtain the corresponding expansions for the angular functions themselves. One extra step is required, however, since the coefficient $a^{\mu}_{\nu,0}(\gamma)$ is itself a function of $\gamma$ and so must be expanded as well. This is readily accomplished using the relation

$$a^{\mu}_{\nu,0}(\gamma) = \left( \sum_{k=-\infty}^{\infty} \left( \frac{a^{\mu}_{\nu,k}(\gamma)}{a^{\mu}_{\nu,0}(\gamma)} \right)^2 \frac{2\,\nu+1}{2\,\nu+2\,k+1} \, \frac{(\nu+\mu+1)_k}{(\nu-\mu+1)_k} \right)^{-1/2}.$$

For large $\gamma$, asymptotic series expansions are also known for the spheroidal wave functions. For integer $m$, $n$ the angular functions reduce to Hermite/Laguerre polynomials as $\gamma \to \infty / \gamma \to \mathrm{i}\,\infty$, and asymptotic expansions in descending powers of $\gamma$ can be found [2]. In our package, these asymptotic expansions are computed using the method just described for the power series.

## IV. Results and Discussion

The most comprehensive sources of tabulated values of spheroidal functions are Flammer [2], Stratton *et al.* [3] and Van Buren *et al.* [19]. Some of the tables from Flammer are reproduced in Tables 21.1-4 of Abramowitz and Stegun [10]. Some minor errors in these works have already been pointed out by Li *et al.* [9]. We have compared the output of our package with all of these sources, and found agreement up to the precision to which the tabulated values are given.

With the availability of packages such as the one we have developed, there is clearly little need for exhaustive tabulations of numerical values. However, it is useful to give some high-precision numerical values to provide a benchmark for comparison with other programs. In Appendix D we provide a set of such values. They are presented to 25 digits and are intended only to represent an illustrative sample.

Because no tables or software packages presently available are capable of generating results to arbitrary precision, the most reliable way to check the validity of our numerical functions is to perform self-consistency tests. Although there are relatively few analytic results available for the spheroidal wave functions, there are several important tests (each of which we have applied to our package with perfect results):

- *Exact solutions*: for $n = 1, 2, \ldots$ we have $\lambda_n^1(n\pi/2) = 0$ [2]. Also, for $\mu = 1/2$, the spheroidal functions reduce to the Mathieu functions ([10], Ch. 20), and the eigenvalues are related to the Mathieu characteristic values $a_r(q)$ by $\lambda_\nu^{1/2}(\gamma) = a_{\nu+1/2}(\gamma^2/4) - \gamma^2/2 - 1/4$ [12]. We can therefore compare the numerical eigenvalues generated by our package to the built-in Mathieu functions in *Mathematica*.

- *Wronskian*: for the spheroidal functions the Wronskian is proportional to $(z^2 - 1)^{-1}$, and for the radial functions it is equal to $(\gamma(z^2 - 1))^{-1}$. This test is the most generally useful, since it can be used for all parameter values.

- *Substitution into the differential equation*: it is straightforward to simply substitute the functions back into the spheroidal differential equation (1) and verify that it is satisfied to the precision of the numerical functions. However, because the second derivative has to be computed numerically, this is not very convenient for testing to a very high level of precision.

- Lastly, a simple check that can always be performed is to generate a certain function value to two different levels of precision (for example 50 and 100 digits). If the two results do not agree up to the precision of the least precise of the two, this indicates there is a problem in the method of calculation. If they do agree, however, it does not guarantee anything—since, for example, the truncated series from which the function is being computed may contain too few terms—but it is a useful guide.

Our numerical package, **Spheroidal.m**, is available online at the URL www.physics.uwa.edu.au/~falloon/spheroidal/spheroidal.html, along with further documentation regarding its use.

## Acknowledgments

PEF is grateful for the support of University Postgraduate Award from the University of Western Australia. Michael Trott and Oleg Marichev of Wolfram Research Inc. provided useful information about general issues concerning the numerical implementation of special functions in *Mathematica*.

## Appendix A—Legendre and spherical Bessel functions

In this appendix we give the definitions of the Legendre and spherical Bessel functions used to the define the spheroidal wave functions. The definitions which we use here differ slightly from those found in [10], and are based on the approach taken in [20]. In particular, we define two *Types* of Legendre function: the functions of Type I, $P_\nu^\mu(z)$ and $Q_\nu^\mu(z)$, are equal to those conventionally used on the interval $z \in (-1, 1)$; the functions of Type II, $\mathfrak{P}_\nu^\mu(z)$ and $\mathfrak{Q}_\nu^\mu(z)$, are equal to those conventionally used for $z \notin (-1, 1)$. The practice of using gothic characters $\mathfrak{P}$, $\mathfrak{Q}$ to denote the functions of Type II follows Meixner and Schäfke [1].

### Legendre functions

The Legendre functions of the first kind are defined by:

$$P_\nu^\mu(z) = \frac{1}{\Gamma(1-\mu)} \frac{(1+z)^{\mu/2}}{(1-z)^{\mu/2}} \, {}_2F_1\!\left(-\nu, \nu+1; 1-\mu; \frac{1-z}{2}\right) \quad \text{(Type I)}, \tag{A1}$$

$$\mathfrak{P}_\nu^\mu(z) = \frac{1}{\Gamma(1-\mu)} \frac{(z+1)^{\mu/2}}{(z-1)^{\mu/2}} \, {}_2F_1\!\left(-\nu, \nu+1; 1-\mu; \frac{1-z}{2}\right) \quad \text{(Type II)}, \tag{A2}$$

where $\Gamma(z)$ is the gamma function ([10], Ch.6) and ${}_2F_1(a, b; c; z)$ is the Gaussian hypergeometric function ([10], Ch.15). Note that these two definitions differ only in their phase, and are trivially related:

$$\mathfrak{P}_\nu^\mu(z) = \frac{(1-z)^{\mu/2}}{(z-1)^{\mu/2}} P_\nu^\mu(z). \tag{A3}$$

For noninteger $\mu$, the functions of the second kind are defined by

$$Q_\nu^\mu(z) = \frac{\pi \csc(\mu\pi)}{2} \left(\cos(\mu\pi) P_\nu^\mu(z) - (\nu - \mu + 1)_{2\mu} P_\nu^{-\mu}(z)\right) \text{ (Type I)}, \tag{A4}$$

$$\mathfrak{Q}_\nu^\mu(z) = \frac{\pi \csc(\mu\pi)}{2} e^{i\mu\pi} \left(\mathfrak{P}_\nu^\mu(z) - (\nu - \mu + 1)_{2\mu} \mathfrak{P}_\nu^{-\mu}(z)\right) \text{ (Type II)}. \tag{A5}$$

The Type I and II functions of the second kind are related by:

$$\mathfrak{Q}_\nu^\mu(z) = e^{i\mu\pi} \frac{(z-1)^{\mu/2}}{(1-z)^{\mu/2}} \left(Q_\nu^\mu(z) + \frac{\pi \csc(\mu\pi)}{2} \left(\frac{(1-z)^\mu}{(z-1)^\mu} - \cos(\mu\pi)\right) P_\nu^\mu(z)\right), \quad \mu \notin \mathbb{Z}, \tag{A6}$$

$$\mathfrak{Q}_\nu^m(z) = (-1)^m \frac{(z-1)^{m/2}}{(1-z)^{m/2}} \left(Q_\nu^m(z) + \frac{\pi}{2} \frac{\sqrt{1-z}}{\sqrt{z-1}} P_\nu^m(z)\right), \quad m \in \mathbb{Z}. \tag{A7}$$

The Type II functions have the following important hypergeometric representation:

$$\begin{aligned}\mathfrak{Q}_\nu^\mu(z) = {} & 2^{-\nu-1} e^{i\mu\pi} \sqrt{\pi} \frac{\Gamma(\mu+\nu+1)}{\Gamma(\nu+3/2)} \\ & \times z^{-\mu-\nu-1} (z+1)^{\mu/2} (z-1)^{\mu/2} {}_2F_1\left(\frac{\mu+\nu+1}{2}, \frac{\mu+\nu}{2} + 1; \nu + \frac{3}{2}; \frac{1}{z^2}\right).\end{aligned} \tag{A8}$$

Because of the factor $\Gamma(\mu+\nu+1)$, this expansion diverges when $\mu + \nu = -1, -2, \ldots$ and hence

$$|Q_\nu^\mu(z)|, |\mathfrak{Q}_\nu^\mu(z)| \to \infty, \quad \text{for } \mu + \nu = -1, -2, \ldots \tag{A9}$$

The following relations for $\nu \to -\nu - 1$ are used in Section 2.2:

$$Q_{-\nu-1}^\mu(z) = \csc(\pi(\mu-\nu)) (\pi \cos(\mu\pi) \cos(\nu\pi) P_\nu^\mu(z) - \sin((\mu+\nu)\pi) Q_\nu^\mu(z)), \tag{A10}$$

$$\mathfrak{Q}_{-\nu-1}^\mu(z) = \csc(\pi(\mu-\nu)) (\pi e^{i\mu\pi} \cos(\nu\pi) \mathfrak{P}_\nu^\mu(z) - \sin((\mu+\nu)\pi) \mathfrak{Q}_\nu^\mu(z)). \tag{A11}$$

In the *Mathematica* system, the Legendre functions of Type I and II are implemented as “type 2” and “type 3” functions (“type 1” is a redundant variant of “type 2” which is only defined for $|z| \leq 1$):

$P_\nu^\mu(z) \longleftrightarrow$ `LegendreP[ν, μ, 2, z]`
$Q_\nu^\mu(z) \longleftrightarrow$ `LegendreQ[ν, μ, 2, z]`
$\mathfrak{P}_\nu^\mu(z) \longleftrightarrow$ `LegendreP[ν, μ, 3, z]`
$\mathfrak{Q}_\nu^\mu(z) \longleftrightarrow$ `LegendreQ[ν, μ, 3, z]`

## Spherical Bessel functions

The spherical Bessel function of the first kind may be defined by

$$j_\nu(z) = \frac{\sqrt{\pi}}{2} \sum_{k=0}^{\infty} \frac{(-1)^k}{\Gamma(k+\nu+3/2)\, k!} \left(\frac{z}{2}\right)^{2k+\nu}. \tag{A12}$$

The function of the second kind is defined by

$$y_\nu(z) = -\sec(\nu\pi)\,(\sin(\nu\pi)\, j_\nu(z) + j_{-\nu-1}(z)). \tag{A13}$$

These functions are not implemented directly in the *Mathematica* system, so we compute them using their relation to the regular Bessel functions $J_\nu(z)$ and $Y_\nu(z)$ ([10], Ch. 9):

$j_\nu(z) \longleftrightarrow$ `Sqrt[π/2] BesselJ[ν + 1/2, z] / Sqrt[z]`

$y_\nu(z) \longleftrightarrow$ `Sqrt[π/2] BesselY[ν + 1/2, z] / Sqrt[z]`

## Appendix B — Flammer's spheroidal functions

In this appendix we give the essential relations between the spheroidal functions of Flammer [2] and those of Meixner [1]. Note that Flammer's functions are only defined for integer parameters $m$, $n$ with $n \geq m \geq 0$.

Eigenvalues:

$$\lambda_{mn}(c) = \lambda_n^m(c) + c^2. \tag{B1}$$

Angular functions:

$$S_{mn}(c, \eta) = \omega_{mn}(c)\,\mathrm{ps}_n^m(\eta; c), \tag{B2}$$

$$S_{mn}^{(2)}(c, \eta) = \omega_{mn}(c)\,\mathrm{qs}_n^m(\eta; c), \tag{B3}$$

where

$$\omega_{mn}(c) = \begin{cases} \dfrac{(-1)^{\frac{n-m}{2}}\,(m+n)!}{2^n\,(\frac{n-m}{2})!\,(\frac{m+n}{2})!}\,\dfrac{1}{\mathrm{ps}_n^m(0; c)}, & n-m\ \text{even}, \\[2ex] \dfrac{(-1)^{\frac{n-m-1}{2}}\,(m+n+1)!}{2^n\,(\frac{n-m-1}{2})!\,(\frac{m+n+1}{2})!}\,\dfrac{1}{\mathrm{ps}_n^{m\prime}(0; c)}, & n-m\ \text{odd}. \end{cases} \tag{B4}$$

Radial functions:

$$R_{mn}^{(1,2)}(c, \xi) = S_n^{m(1,2)}(\xi; c). \tag{B5}$$

## Appendix C—Symmetry relations for the spheroidal functions

The spheroidal wave functions satisfy a number of useful mathematical identities, which they "inherit" from properties the Legendre and spherical Bessel functions. Some of these were originally derived in a number of papers by Meixner (almost all of whose work was published in German—see [1] and references therein), but do not appear in the most popular references on spheroidal wave functions (*e.g.* [2, 10]), and are therefore effectively unavailable to the majority of contemporary readers. The situation is further hampered by the different notations and normalizations in existence: deriving identities valid for Flammer's functions from those in Meixner's is nontrivial. In this appendix we therefore present a concise summary of these identities for the spheroidal wave functions, including all special cases which are not trivially obtained from the general ones. All relations are valid throughout the complex plane, and in particular along branch cuts. A more detailed discussion of the derivation of these identities can be found in [12].

### The transformation $\nu \to -\nu - 1$

Eigenvalues:

$$\lambda^{\mu}_{-\nu-1}(\gamma) = \lambda^{\mu}_{\nu}(\gamma). \tag{C1}$$

Radial normalization factor:

$$A^{\mu}_{-\nu-1}(\gamma) = A^{\mu}_{\nu}(\gamma). \tag{C2}$$

Angular functions, general $\mu$:

$$f^{\mu}_{-\nu-1}(z;\gamma) = f^{\mu}_{\nu}(z;\gamma), \quad f = \mathrm{Ps}, \mathrm{ps}, \tag{C3}$$

$$\mathrm{qs}^{\mu}_{-\nu-1}(z;\gamma) = \csc((\mu-\nu)\pi)\,(\pi\cos(\mu\pi)\cos(\nu\pi)\,\mathrm{ps}^{\mu}_{\nu}(z;\gamma) - \sin((\mu+\nu)\pi)\,\mathrm{qs}^{\mu}_{\nu}(z;\gamma)), \tag{C4}$$

$$\mathrm{Qs}^{\mu}_{-\nu-1}(z;\gamma) = \csc((\mu-\nu)\pi)\,(\pi\, e^{\mathrm{i}\mu\pi}\cos(\nu\pi)\,\mathrm{Ps}^{\mu}_{\nu}(z;\gamma) - \sin((\mu+\nu)\pi)\,\mathrm{Qs}^{\mu}_{\nu}(z;\gamma)). \tag{C5}$$

Angular functions of the second kind, integer $m$:

$$\mathrm{qs}^{m}_{-\nu-1}(z;\gamma) = \mathrm{qs}^{\mu}_{\nu}(z;\gamma) - \pi\cot(\nu\pi)\,\mathrm{ps}^{\mu}_{\nu}(z;\gamma), \tag{C6}$$

$$\mathrm{Qs}^{m}_{-\nu-1}(z;\gamma) = \mathrm{Qs}^{\mu}_{\nu}(z;\gamma) - \pi\cot(\nu\pi)\,\mathrm{Ps}^{\mu}_{\nu}(z;\gamma). \tag{C7}$$

Radial functions, general $\nu$:

$$S^{\mu\,(1)}_{-\nu-1}(z;\gamma) = -\sin(\nu\pi)\,S^{\mu(1)}_{\nu}(z;\gamma) - \cos(\nu\pi)\,S^{\mu(2)}_{\nu}(z;\gamma), \tag{C8}$$

$$S^{\mu\,(2)}_{-\nu-1}(z;\gamma) = \cos(\nu\pi)\,S^{\mu(1)}_{\nu}(z;\gamma) - \sin(\nu\pi)\,S^{\mu(2)}_{\nu}(z;\gamma). \tag{C9}$$

Radial functions, integer $n$:

$$S^{\mu}_{-n-1}{}^{(1)}(z;\gamma) = (-1)^{n+1}\, S^{\mu\,(2)}_{n}(z;\gamma), \tag{C10}$$

$$S^{\mu}_{-n-1}{}^{(2)}(z;\gamma) = (-1)^{n}\, S^{\mu\,(1)}_{n}(z;\gamma). \tag{C11}$$

## The transformation $\mu \to -\mu$

Eigenvalues:

$$\lambda^{-\mu}_{\nu}(\gamma) = \lambda^{\mu}_{\nu}(\gamma). \tag{C12}$$

Angular functions, general $\mu$:

$$\mathrm{ps}^{-\mu}_{\nu}(z;\gamma) = \frac{\Gamma(\nu-\mu+1)}{\Gamma(\nu+\mu+1)}\Big(\cos(\mu\,\pi)\,\mathrm{ps}^{\mu}_{\nu}(z;\gamma) - \frac{2}{\pi}\sin(\mu\,\pi)\,\mathrm{qs}^{\mu}_{\nu}(z;\gamma)\Big), \tag{C13}$$

$$\mathrm{qs}^{-\mu}_{\nu}(z;\gamma) = \frac{\Gamma(\nu-\mu+1)}{\Gamma(\nu+\mu+1)}\Big(\cos(\mu\,\pi)\,\mathrm{qs}^{\mu}_{\nu}(z;\gamma) + \frac{\pi}{2}\sin(\mu\,\pi)\,\mathrm{ps}^{\mu}_{\nu}(z;\gamma)\Big), \tag{C14}$$

$$\mathrm{Ps}^{-\mu}_{\nu}(z;\gamma) = \frac{\Gamma(\nu-\mu+1)}{\Gamma(\nu+\mu+1)}\Big(\mathrm{Ps}^{\mu}_{\nu}(z;\gamma) - \frac{2}{\pi}\,e^{-\mathrm{i}\,\mu\,\pi}\sin(\mu\,\pi)\,\mathrm{Qs}^{\mu}_{\nu}(z;\gamma)\Big), \tag{C15}$$

$$\mathrm{Qs}^{-\mu}_{\nu}(z;\gamma) = \frac{\Gamma(\nu-\mu+1)}{\Gamma(\nu+\mu+1)}\,e^{-2\,\mathrm{i}\,\mu\,\pi}\,\mathrm{Qs}^{\mu}_{\nu}(z;\gamma). \tag{C16}$$

Angular functions, integer $m$:

$$f^{-m}_{\nu}(z;\gamma) = (-1)^{m}\,\frac{\Gamma(\nu-m+1)}{\Gamma(\nu+m+1)}\,f^{m}_{\nu}(z;\gamma), \quad f = \mathrm{ps},\,\mathrm{qs}, \tag{C17}$$

$$f^{-m}_{\nu}(z;\gamma) = \frac{\Gamma(\nu-m+1)}{\Gamma(\nu+m+1)}\,f^{m}_{\nu}(z;\gamma), \quad f = \mathrm{Ps},\,\mathrm{Qs}. \tag{C18}$$

Radial functions:

$$S^{-\mu\,(k)}_{\nu}(z;\gamma) = S^{\mu\,(k)}_{\nu}(z;\gamma), \quad k = 1,\,2,\,3,\,4. \tag{C19}$$

## The transformation $z \to -z$

Angular functions of Type I, general $\mu$, $\nu$:

$$\mathrm{ps}^{\mu}_{\nu}(-z;\gamma) = \cos((\mu+\nu)\,\pi)\,\mathrm{ps}^{\mu}_{\nu}(z;\gamma) - \frac{2}{\pi}\sin((\mu+\nu)\,\pi)\,\mathrm{qs}^{\mu}_{\nu}(z;\gamma), \tag{C20}$$

$$\mathrm{qs}^{\mu}_{\nu}(-z;\gamma) = -\cos((\mu+\nu)\,\pi)\,\mathrm{qs}^{\mu}_{\nu}(z;\gamma) - \frac{\pi}{2}\sin((\mu+\nu)\,\pi)\,\mathrm{ps}^{\mu}_{\nu}(z;\gamma). \tag{C21}$$

Angular functions of Type II, general $\mu$, $\nu$ and $z \notin (-1,\,1)$:

$$\mathrm{Ps}_\nu^\mu(-z;\gamma) = \exp\!\left(\pi\,\nu\,\sqrt{-z^2}\;\big/ z\right)\mathrm{Ps}_\nu^\mu(z;\gamma) - \frac{2}{\pi}\,e^{-\mathrm{i}\,\mu\,\pi}\,\sin(\pi\,(\mu+\nu))\,\mathrm{Qs}_\nu^\mu(z;\gamma), \tag{C22}$$

$$\mathrm{Qs}_\nu^\mu(-z;\gamma) = -\exp\!\left(-\pi\,\nu\,\sqrt{-z^2}\;\big/ z\right)\mathrm{Qs}_\nu^\mu(z;\gamma). \tag{C23}$$

Angular functions, integer $m$, $n$:

$$\mathrm{ps}_n^m(-z;\gamma) = (-1)^{m+n}\,\mathrm{ps}_n^m(z;\gamma), \tag{C24}$$

$$\mathrm{qs}_n^m(-z;\gamma) = (-1)^{m+n+1}\,\mathrm{qs}_n^m(z;\gamma), \tag{C25}$$

$$\mathrm{Ps}_n^m(-z;\gamma) = (-1)^n\,\mathrm{Ps}_n^m(z;\gamma),\quad z\notin(-1,\,1), \tag{C26}$$

$$\mathrm{Qs}_\nu^\mu(-z;\gamma) = (-1)^{n+1}\,\mathrm{Qs}_\nu^\mu(z;\gamma),\quad z\notin(-1,\,1). \tag{C27}$$

Radial functions, general $\nu$:

$$S_\nu^{\mu\,(1)}(-z;\gamma) = (-\gamma\,z)^\nu\,(\gamma\,z)^{-\nu}\,S_\nu^{\mu\,(1)}(z;\gamma), \tag{C28}$$

$$\begin{aligned} &S_\nu^{\mu\,(2)}(-z;\gamma) = \\ &\quad -(-\gamma\,z)^{-\nu}\,(\gamma\,z)^\nu\,(S_\nu^{\mu\,(2)}(z;\gamma) + (1+(-\gamma\,z)^{2\,\nu}\,(\gamma\,z)^{-2\,\nu})\tan(\nu\,\pi)\,S_\nu^{\mu\,(1)}(z;\gamma)). \end{aligned} \tag{C29}$$

Radial functions, integer $n$:

$$S_n^{\mu\,(1)}(-z;\gamma) = (-1)^n\,S_n^{\mu\,(1)}(z;\gamma), \tag{C30}$$

$$S_n^{\mu\,(2)}(-z;\gamma) = (-1)^{n+1}\,S_n^{\mu\,(2)}(z;\gamma). \tag{C31}$$

### Relations between angular functions of Type I and II

$$\mathrm{Ps}_\nu^\mu(z;\gamma) = \frac{(1-z)^{\mu/2}}{(z-1)^{\mu/2}}\,\mathrm{ps}_\nu^\mu(z;\gamma), \tag{C32}$$

$$\mathrm{Qs}_\nu^\mu(z;\gamma) = e^{\mathrm{i}\,\mu\,\pi}\,\frac{(z-1)^{\mu/2}}{(1-z)^{\mu/2}}\left(\mathrm{qs}_\nu^\mu(z;\gamma) + \frac{\pi\csc(\mu\,\pi)}{2}\left(\frac{(1-z)^\mu}{(z-1)^\mu} - \cos(\mu\,\pi)\right)\mathrm{ps}_\nu^\mu(z;\gamma)\right). \tag{C33}$$

The second of these must be treated specially for integer $m$:

$$\mathrm{Qs}_\nu^m(z;\gamma) = (-1)^m\,\frac{(z-1)^{m/2}}{(1-z)^{m/2}}\left(\mathrm{qs}_\nu^m(z;\gamma) + \frac{\pi}{2}\,\frac{\sqrt{1-z}}{\sqrt{z-1}}\,\mathrm{ps}_\nu^m(z;\gamma)\right). \tag{C34}$$

### Relations between angular and radial functions

In this section we present the set of relations between the angular and radial functions which we use to compute the functions throughout the complex $z$-plane. They can be derived using Eq. (19) and the identities given in this appendix. The forms presented here, which are completely general and valid along branch cuts, have not previously appeared in the literature.

Radial functions, general $\mu$, $\nu$:

$$S_\nu^{\mu\,(1)}(z;\gamma) = K_\nu^\mu(\gamma)\,\frac{\sin((\mu-\nu)\,\pi)}{\pi}\,e^{-\mathrm{i}\,(\mu+\nu)\,\pi}\,\frac{(1-1/z^2)^{\mu/2}\,(\gamma\,z)^\nu}{\gamma^\nu\,z^{\nu-\mu}\,(z-1)^{\mu/2}\,(z+1)^{\mu/2}}\,\mathrm{Qs}_{-\nu-1}^\mu(z;\gamma), \tag{C35}$$

$$S_\nu^{\mu\,(2)}(z;\gamma) = \sec(\nu\,\pi)\left(S_{-\nu-1}^{\mu}{}^{(1)}(z;\gamma) - \sin(\nu\,\pi)\,S_\nu^{\mu\,(1)}(z;\gamma)\right). \tag{C36}$$

Radial functions, integer $m$, $n$:

$$S_n^{m\,(1)}(z;\gamma) = K_n^m(\gamma)\,\frac{(1-1/z^2)^{m/2}\,z^m}{(z-1)^{m/2}\,(z+1)^{m/2}}\,\mathrm{Ps}_n^m(z;\gamma), \tag{C37}$$

$$S_n^{m\,(2)}(z;\gamma) = \frac{(-1)^{m+1}}{\gamma\,K_n^{-m}(\gamma)\,A_n^m(\gamma)\,A_n^{-m}(\gamma)}\,\frac{(1-1/z^2)^{m/2}\,z^m}{(z-1)^{m/2}\,(z+1)^{m/2}}\,\mathrm{Qs}_n^m(z;\gamma). \tag{C38}$$

Type II angular functions, general $\mu$, $\nu$:

$$\mathrm{Qs}_\nu^\mu(z;\gamma) = \frac{\pi\csc((\mu+\nu)\,\pi)\,e^{\mathrm{i}\,(\mu+\nu)\,\pi}}{K_{-\nu-1}^\mu(\gamma)}\,\frac{(z-1)^{\mu/2}\,(z+1)^{\mu/2}\,(\gamma\,z)^{\nu+1}}{(1-1/z^2)^{\mu/2}\,\gamma^{\nu+1}\,z^{\mu+\nu+1}}\,S_{-\nu-1}^{\mu}{}^{(1)}(z;\gamma), \tag{C39}$$

$$\mathrm{Ps}_\nu^\mu(z;\gamma) = \frac{\sec(\nu\,\pi)}{\pi}\,e^{-i\,\mu\,\pi}\,(\sin(\pi\,(\mu+\nu))\,\mathrm{Qs}_\nu^\mu(z;\gamma) - \sin((\nu-\mu)\,\pi)\,\mathrm{Qs}_{-\nu-1}^\mu(z;\gamma)). \tag{C40}$$

Type II angular functions, integer $m$, $n$:

$$\mathrm{Ps}_n^m(z;\gamma) = \frac{1}{K_n^m(\gamma)}\,\frac{(1-1/z^2)^{m/2}\,z^m}{(z-1)^{m/2}\,(z+1)^{m/2}}\,S_n^{m\,(1)}(z;\gamma), \tag{C41}$$

$$\mathrm{Qs}_n^m(z;\gamma) = (-1)^{m+1}\,\gamma\,K_n^{-m}(\gamma)\,A_n^m(\gamma)\,A_n^{-m}(\gamma)\,\frac{(z-1)^{m/2}\,(z+1)^{m/2}}{(1-1/z^2)^{m/2}\,z^m}\,S_n^{m\,(2)}(z;\gamma). \tag{C42}$$

Relations involving the Type I angular functions, $\mathrm{ps}_\nu^\mu(z;\gamma)$ and $\mathrm{qs}_\nu^\mu(z;\gamma)$, can be obtained from the last four equations by using Eqs. (C32-34).

## Appendix D—Tables of numerical values

In this appendix we present a set of numerical values, to 25 digits of precision, for all of Meixner's spheroidal functions. The intention is to provide enough values and to high enough precision to facilitate comparison with any future software implementation. Tables 1 and 2 contain the eigenvalues $\lambda_\nu^\mu(\gamma)$ for integer and complex parameters respectively; Table 3 contains the joining and normalization factors $K_\nu^\mu(\gamma)$ and $A_\nu^\mu(\gamma)$; finally, Tables 4 and 5 contain the angular and radial functions and their derivatives. Values for the Type II angular functions, as well as all of Flammer's functions, are not given, since they can easily be found from those presented here.

**Table 1** Eigenvalues $\lambda_n^m(\gamma)$ for integer $m$, $n$ and real $\gamma^2$.

| $m$ | $n$ | $\gamma$ | $\lambda_n^m(\gamma)$ | $\lambda_n^m(\mathrm{i}\,\gamma)$ |
|---|---|---|---|---|
| 0 | 0 | 10 | −90.77169 57027 50054 84898 77312 | 18.97205 60550 42243 81391 09191 |
| | | 100 | −9900.75189 88910 16747 44954 21523 | 198.99747 46340 82548 13572 48103 |
| | 1 | 10 | −71.86653 62671 73272 18538 10250 | 18.97206 19762 54415 92684 71575 |
| | | 100 | −9701.75954 33440 82366 62256 40610 | 198.99747 46340 82548 13572 48103 |
| 1 | 1 | 10 | −89.71223 12326 08531 82924 20084 | 37.88064 98956 19453 22628 71049 |
| | | 100 | −9899.74682 23865 85061 62347 24355 | 397.98984 67939 13121 45974 40125 |
| | 2 | 10 | −70.66108 19583 85518 52994 19784 | 37.88084 87977 73011 20481 64244 |
| | | 100 | −9700.74415 65958 58817 37915 37426 | 397.98984 67939 13121 45974 40125 |

**Table 2** Eigenvalues $\lambda_\nu^\mu(\gamma)$ for complex $\mu$, $\nu$, $\gamma$. Here we use the abbrevation $\alpha = 1 + \mathrm{i}$.

| $\mu$ | $\nu$ | $c$ | $\mathrm{Re}(\lambda_\nu^\mu(\gamma))$ | $\mathrm{Im}(\lambda_\nu^\mu(\gamma))$ |
|---|---|---|---|---|
| 0 | 0 | $\alpha$ | 0.05947 27697 35031 26247 06156 | −1.33717 48778 05399 97103 72379 |
| | | $10\,\alpha$ | 9.24076 62146 34603 35159 57443 | −189.98934 85956 57553 67515 08696 |
| 0 | $\alpha$ | 1 | 0.50186 77624 67004 53075 16267 | 2.95073 69925 18211 20706 17898 |
| | $10\,\alpha$ | | 9.50003 16512 34204 66677 88169 | 209.99925 73181 59354 50064 18858 |
| $\alpha$ | 0 | 1 | −0.90781 92346 93494 49431 33571 | 0.93747 61281 94795 84236 49580 |
| $10\,\alpha$ | | | −13.78249 20414 53639 96320 69793 | 17.03738 91416 68651 13441 81798 |
| $\alpha$ | $\alpha$ | $\alpha$ | 1.14617 35587 36254 25050 29932 | 1.33182 58434 94567 67063 46083 |
| $10\,\alpha$ | $10\,\alpha$ | $10\,\alpha$ | 13.77544 66537 42879 55398 69300 | 14.13344 43105 19156 64488 99153 |

**Table 3** Joining factor, $K_n^m(\gamma)$, and radial normalization factor $A_n^m(\gamma)$

| $m$ | $n$ | $\gamma$ | $K_n^m(\gamma)$ | $A_n^m(\gamma)$ |
|---|---|---|---|---|
| 0 | 0 | 10 | 428.00699 32832 74587 46082 31771 | 0.00092 59959 00168 65734 97377 |
| | | 10 i | 129.95559 15348 21883 48851 72999 | 4.35228 56879 68459 42426 84086 |
| | 1 | 10 | 89.19337 15984 63365 35496 77126 | 0.00444 35150 58595 83160 08489 |
| | | 10 i | 225.08939 49188 27718 78315 26217 i | 2.51279 49340 42137 95801 16552 |
| 1 | 1 | 10 | 494.80660 08906 73485 53502 72884 | −0.00008 45665 54309 12744 77142 |
| | | 10 i | 27.46214 85726 35368 05266 46286 i | 0.55989 62979 48596 23342 04584 |
| | 2 | 10 | 43.80698 90817 46872 38955 24520 | −0.00107 62847 63641 61970 54026 |
| | | 10 i | −20.46826 39377 38829 05641 78446 | 0.75119 25147 80086 50861 25805 |

**Table 4** Type I angular functions and their derivatives at $z = 0$.

| $m$ | $n$ | $\gamma$ | $\mathrm{ps}_n^m\,(0;\gamma)$ | $\mathrm{ps}_{n+1}^m{}'\,(0;\gamma)$ |
|---|---|---|---|---|
| 0 | 0 | 10 | $1.86950\ 13198\ 83220\ 32378\ 66070 \times 10^{0}$ | $4.62218\ 68979\ 44534\ 31859\ 57783 \times 10^{0}$ |
| | | 10i | $8.13921\ 06153\ 91477\ 31355\ 92685 \times 10^{-4}$ | $4.20017\ 80506\ 23196\ 12220\ 71385 \times 10^{-3}$ |
| 1 | 1 | 10 | $-1.52903\ 37582\ 54318\ 09757\ 33869 \times 10^{0}$ | $-8.82749\ 07181\ 87103\ 21096\ 49776 \times 10^{0}$ |
| | | 10i | $-4.10717\ 23604\ 57252\ 74666\ 32257 \times 10^{-3}$ | $-4.33152\ 86911\ 29750\ 60250\ 68055 \times 10^{-2}$ |
| $m$ | $n$ | $\gamma$ | $\mathrm{qs}_{n+1}^m(0;\gamma)$ | $\mathrm{qs}_n^m{}'(0;\gamma)$ |
| 0 | 0 | 10 | $-4.27174\ 98257\ 69345\ 65574\ 94192 \times 10^{-6}$ | $4.58661\ 56819\ 97616\ 23157\ 52064 \times 10^{-7}$ |
| | | 10i | $-1.50330\ 25515\ 69445\ 99770\ 03079 \times 10^{3}$ | $2.32730\ 07180\ 66502\ 65903\ 60560 \times 10^{4}$ |
| 1 | 1 | 10 | $-9.27251\ 18702\ 47298\ 25165\ 14180 \times 10^{-6}$ | $4.19596\ 49830\ 13982\ 19782\ 26061 \times 10^{-7}$ |
| | | 10i | $4.93493\ 01484\ 86571\ 35347\ 97135 \times 10^{2}$ | $-2.89127\ 68871\ 67718\ 83087\ 43378 \times 10^{3}$ |

**Table 5** Radial functions and their derivatives at $z = 1.005$.

| $m$ | $n$ | $\gamma$ | $S_n^{m(1)}(z;\gamma)$ | $S_n^{m(1)\prime}(z;\gamma)$ |
|---|---|---|---|---|
| 2 | 2 | 1 | $6.61191\ 32248\ 51537\ 44227\ 25009 \times 10^{-4}$ | $1.32472\ 88100\ 07683\ 20705\ 27852 \times 10^{-1}$ |
| 2 | 2 | 2 | $2.56592\ 96586\ 98996\ 40081\ 40566 \times 10^{-3}$ | $5.12978\ 72006\ 11894\ 29814\ 83008 \times 10^{-1}$ |
| 2 | 3 | 3 | $2.20653\ 45978\ 82418\ 05038\ 85691 \times 10^{-3}$ | $4.42319\ 54640\ 28593\ 94205\ 30600 \times 10^{-1}$ |
| 2 | 3 | 4 | $4.68276\ 42681\ 95501\ 75619\ 52436 \times 10^{-3}$ | $9.34757\ 21512\ 11403\ 78681\ 71462 \times 10^{-1}$ |
| $m$ | $n$ | $\gamma$ | $S_n^{m(2)}(z;\gamma)$ | $S_n^{m(2)\prime}(z;\gamma)$ |
| 2 | 2 | 1 | $-3.74977\ 22396\ 54243\ 54812\ 78539 \times 10^{2}$ | $7.57364\ 90437\ 91073\ 13553\ 02702 \times 10^{4}$ |
| 2 | 2 | 2 | $-4.85222\ 67972\ 28220\ 36109\ 36955 \times 10^{1}$ | $9.73698\ 58589\ 49359\ 43573\ 03506 \times 10^{3}$ |
| 2 | 3 | 3 | $-3.74287\ 18891\ 97107\ 67822\ 75646 \times 10^{1}$ | $7.56605\ 12493\ 58967\ 24757\ 30118 \times 10^{3}$ |
| 2 | 3 | 4 | $-1.33399\ 79013\ 10628\ 13090\ 07387 \times 10^{1}$ | $2.66253\ 29643\ 35609\ 64101\ 07459 \times 10^{3}$ |